\documentclass[10pt,a4paper]{article}
\usepackage{jheppub_kim}
\usepackage{pdflscape}
\usepackage{amsmath}
\usepackage{amssymb}
\usepackage{dcolumn}
\usepackage{bm}
\usepackage{color}
\usepackage{epsfig}
\usepackage{amsfonts}
\usepackage{graphicx}
\usepackage{subfigure}
\usepackage{dcolumn}
\usepackage{color}

\newcommand{\tcr}{\textcolor{red}}

 \newcommand{\rbm}[1]{{\color{magenta}\bf [Robb: #1]}}

\begin{document}
\title{Note on the thermodynamic stability of a black ring at quantum scales}

\author[a]{Robert B. Mann,}
\author[b]{Behnam Pourhassan,}
\author[c]{Prabir Rudra}

\affiliation[a] {Department of Physics and Astronomy, University of Waterloo, Waterloo, Ontario N2L 3G1, Canada.}
\affiliation[a] {Perimeter Institute for Theoretical Physics, 31 Caroline St. N., Waterloo, Ontario N2L 2Y5, Canada.}

\affiliation[b] {School of Physics, Damghan University, P. O. Box 3671641167, Damghan, Iran.}

\affiliation[c] {Department of Mathematics, Asutosh College, Kolkata-700 026, India.}

\emailAdd{rbmann@uwaterloo.ca}

\emailAdd{b.pourhassan@du.ac.ir}

\emailAdd{prudra.math@gmail.com, rudra@associates.iucaa.in}

\abstract{In this paper, the thermodynamic properties of a thin
black ring in AdS space-time is explored when the size of the ring
is comparable to quantum scales. The angular momentum to mass
ratio of this system has an upper limit, which is the cosmological
radius of the black ring. It is found that the small black ring
will be thermodynamically stable due to the effects introduced by
thermal fluctuations. However, we find that the black ring is less
stable than thermal AdS. Thermodynamic analysis indicates that
there is no critical point, but there is Hawking-Page transition
to radiation, which is confirmed by the Gibbs free energy
analysis.}


\maketitle

\section{Introduction}
Four-dimensional black holes are natural solutions of
Einstein's theory of general relativity (GR). The simplest of these is the
spherically symmetric solution known as the Schwarzschild metric
 \cite{1}, which describes  a static black hole (BH) that has no
electric charge and no angular momentum. The only parameter  that
distinguishes various Schwarzschild BHs is the   mass. The linear
stability of these BHs was investigated by Dotti  \cite{1-1}. Adding an electric charge yields the
 Reissner-Nordstr\"{o}m BH, which is super-radiantly unstable
against spherical perturbations of a charged scalar field
\cite{2}.  Alternatively, adding rotation yields the Kerr metric \cite{3}, which is
an axially-symmetric vacuum solution of Einstein equations describing a BH with  mass and angular
momentum. An interesting parameter here is the ratio ($b$) of angular momentum to
squared mass,  known as the Kerr parameter,  whose
absolute value is smaller than unity ($\mid b\mid<1$).
The most general vacuum solution of the Einstein-Maxwell equations in GR is called the Kerr-Newman metric.
characterized by three parameters:  mass, charge and angular
momentum. This metric describes a charged and rotating BH, whose linear mode is stable within
Einstein-Maxwell theory \cite {4}, though they have super-radiant instabilities \cite{1807.06263}. If a
cosmological constant $\Lambda$ is included \cite{5} then all of these BHs generalize to what are called
anti-de Sitter (AdS) ($\Lambda < 0$)  and de-Sitter (dS) BHs  ($\Lambda > 0$).\\
These four-dimensional BHs can be extended to  higher
dimensional space-times \cite{6}.  However their stability properties can change. For example,
it has been argued by Konoplya and Zhidenko \cite{7} that the
higher-dimensional Reissner-Nordstr\"{o}m-de Sitter BHs are
gravitationally unstable. Along with gravitational stability,
thermodynamic stability of black objects is also an interesting
topic of study \cite{8}.\\
One of the important black objects in higher dimensions is the
black ring \cite{9},  which has horizon topology of $S^{1}\times
S^{d-3}$ in $d$-dimensional space-time. Black rings are classified as either fat or thin,
depending on their shape.  Numerical methods have been used to show that a black ring in five dimensions
is gravitationally unstable \cite{10}.
This instability may be just like those
of black strings and p-branes in higher dimensions \cite{11}, and
numerical evidence that the end state of these instabilities can
violate  the weak cosmic censorship conjecture has been provided \cite{12}.
Fat rings are unstable \cite{10}, which has been
demonstrated using local Penrose inequalities.\\
There is  very little information in literature about the stability of thin
rings. A supersymmetric five-dimensional black ring having an
event horizon of topology $S^{1}\times S^{2}$ has been constructed \cite{13}.
and a certain two-charge supersymmetric
state of such configuration was studied  from
the statistical viewpoint  \cite{13-1}.  The exact microscopic entropy of a black ring
using the M-theory was subsequently obtained \cite{15}, and the
mass-angular-momentum inequality for such black rings has been used
  to gain insight into the standard picture of
gravitational collapse \cite{14}. A string
theoretic description of near extremal black rings was proposed by
Larsen \cite{16}. The thermodynamical properties of a dipole black
ring was investigated by Astefanesei and Radu using the quasilocal
formalism  \cite{17}.\\
In this paper we are interested in investigating the
thermodynamical stability of thin black rings in AdS space that
incorporate expected quantum corrections to the entropy that are
logarithmic in nature in the horizon area \cite{Mann:1997hm}.
Although a black ring in AdS space is gravitationally and
thermodynamically unstable, in the presence of thermal
fluctuations, the scenario is expected to be slightly different at
small scales: the thin black ring in AdS space may be stable due
to thermal fluctuations at small scales due to the quantum
effects. An analysis of the Gibbs free energy indicates  that small black rings with low angular momentum are
thermodynamically stable.\\
We begin by briefly describing the black ring parameters and their
relationship with  classical angular momenta in  Sec. 2. In Sec. 3,
we explore the thermodynamic aspects of the system and  end with a
discussion and conclusions in Sec. 4.

\section{Black Ring}

It is known that $d$-dimensional AdS space-time admits a thin black ring,
constructed from a thin black string of width $r_{0}$ transformed
to a circle of radius $R$, with horizon topology of $S^{1}\times
S^{d-3}$ \cite{18-0}. Beginning with global AdS  space-time
\begin{equation}\label{metric}
ds^{2}=-fd\tau^{2}+\frac{d\rho^{2}}{f}+\rho^{2}(d\theta^{2}+\sin^{2}\theta
d\Omega_{d-4}^{2}+\cos^{2}\theta d\psi^{2}),
\end{equation}
where
\begin{equation}\label{metric f}
f=1+\frac{\rho^{2}}{l^{2}},
\end{equation}
and $d\Omega_{d-4}^{2}$ is the metric of a $(d-4)$-dimensional
unit sphere with volume,
\begin{equation}\label{Volume d-4}
\Omega_{d-4}=\frac{2\pi^{\frac{d-3}{2}}}{\Gamma\left(\frac{d-3}{2}\right)}.
\end{equation}
the
black ring is located at $\rho=R$ on the $\theta=0$ plane, so that $r_{0}\ll R$ and $r_{0}\ll l$, where $l$ is the cosmological radius  \cite{18}.
At large distances  the gravitational field created by the ring (in directions  transverse to the ring)
 is the same as that of an equivalent circular distribution of energy-momentum centered at  $\rho=R$, $\theta=0$.
 Close to this it is possible to choose a set of adapted coordinates in
AdS so that  the radial coordinate $r$ measures the transverse distance away from the circle at  $r = 0$ (the location of the ring), where surfaces of constant $r$ have ring-like topology.  A complete description appears in ref.   \cite{18}.

\begin{figure}
\begin{center}$
\begin{array}{cccc}
\includegraphics[width=55 mm]{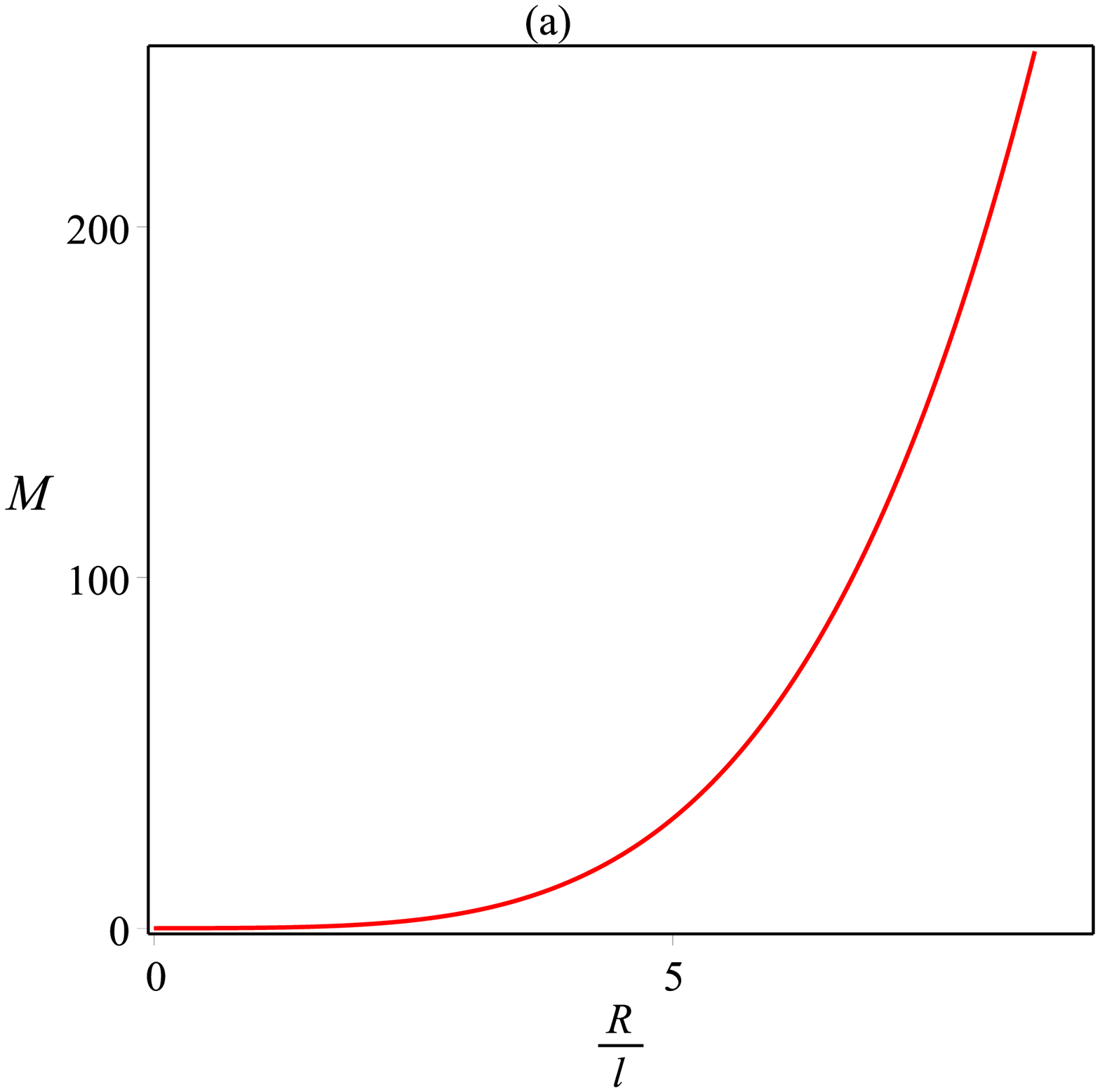}\includegraphics[width=55 mm]{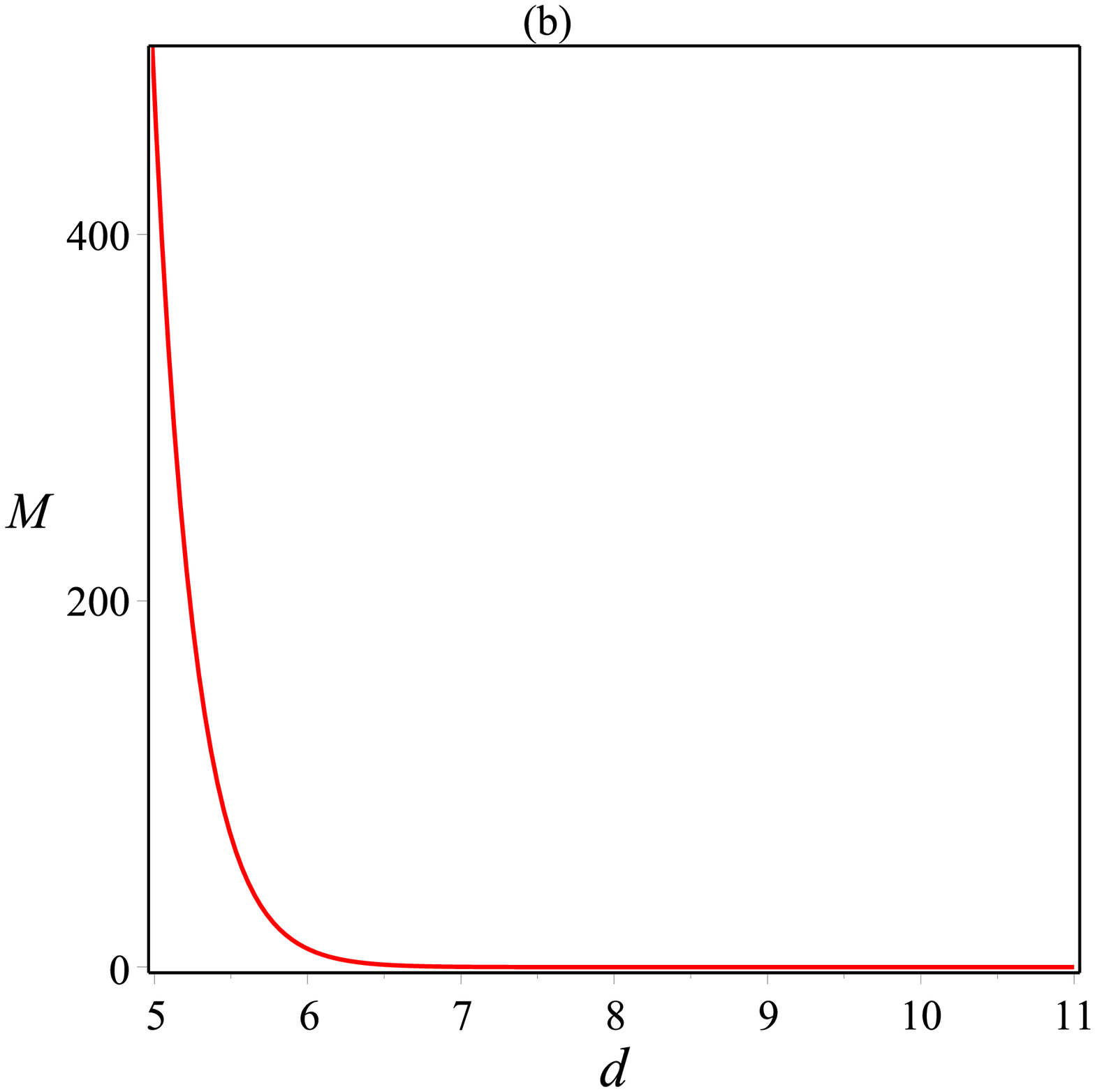}
\end{array}$
\end{center}
\caption{Typical behavior of the mass as a function of $R/l$  for $r_{0}/l=0.01$.
(a) in terms of $R/l$ for $d=5$; (b) in terms of $d$ for $R/l=10$. }
\label{figM}
\end{figure}

\begin{figure}
\begin{center}$
\begin{array}{cccc}
\includegraphics[width=55 mm]{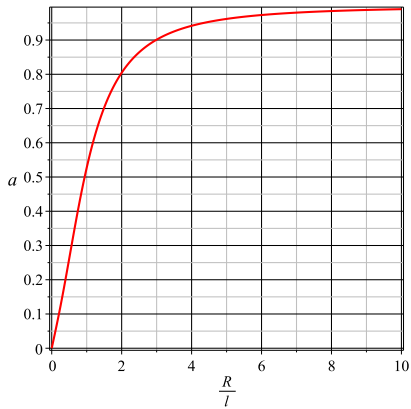}
\end{array}$
\end{center}
\caption{Typical behavior of the values of the angular momentum
per unit mass in terms of $R/l$ for $d = 5$.}
\label{figa}
\end{figure}

The mass of the  ring in units of the Planck mass is
\begin{equation}\label{M}
M=\frac{r_{0}^{d-4}\Omega_{d-3}R(d-2)\left(1+\frac{R^2}{l^2}\right)^{\frac{3}{2}}}{8l_{p}^{d-2}},
\end{equation}
where $\Omega_{d-3}$ obtained via (\ref{Volume d-4}).
Clearly $M$ is an increasing function of $R$, whereas it is a decreasing function of $d$, which we illustrate in Fig. \ref{figM}.
The other conserved quantity associated with the black ring is its angular momentum, given by  \cite{18}
\begin{equation}\label{J}
J=\frac{r_{0}^{d-4}\Omega_{d-3}R^{2}\sqrt{(1+(d-2)\frac{R^{2}}{l^{2}})(d-3+(d-2)\frac{R^{2}}{l^{2}})}}{8l_{p}^{2}}.
\end{equation}
It is clear that the angular momentum is also an increasing function of $R$.  The angular momentum per unit mass \cite{3}
\begin{equation}\label{a}
a\equiv\frac{J}{M}=\frac{R}{d-2}\sqrt{\frac{(1+(d-2)\frac{R^{2}}{l^{2}})(d-3+(d-2)\frac{R^{2}}{l^{2}})}{\left(1+\frac{R^{2}}{l^{2}}\right)^{3}}}\leq l
\end{equation}
becomes a $d$-independent constant at large values of $R$, whose upper bound is always given by $l$;
 $a\rightarrow l$ for $R/l \gg 1$. We illustrate this  in Fig. \ref{figa}  for $d=5$;  other dimensions also yield
qualitatively similar results. Note from \eqref{a} that
$a$ is independent of $r_{0}$.
The angular velocity of the horizon
\begin{equation}\label{Omega}
\Omega=\sqrt{\frac{(1+\frac{R^{2}}{l^{2}})(1+(d-2)\frac{R^{2}}{l^{2}})}{R^{2}(d-3+(d-2)\frac{R^{2}}{l^{2}})}}
\end{equation}
is the thermodynamic conjugate of $J$, and is a decreasing function of  $R$.

\begin{figure}
\begin{center}$
\begin{array}{cccc}
\includegraphics[width=55 mm]{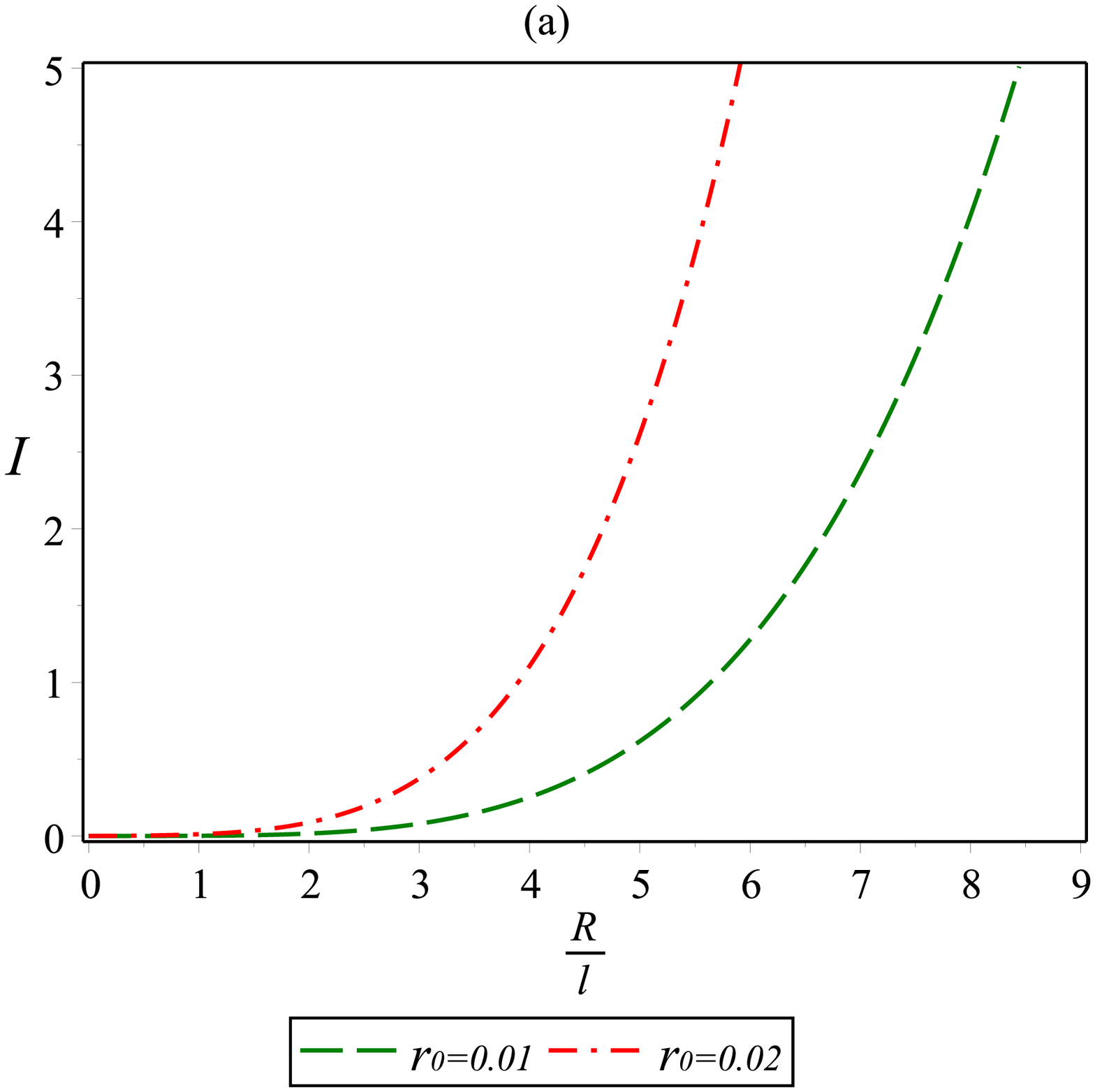}\includegraphics[width=55 mm]{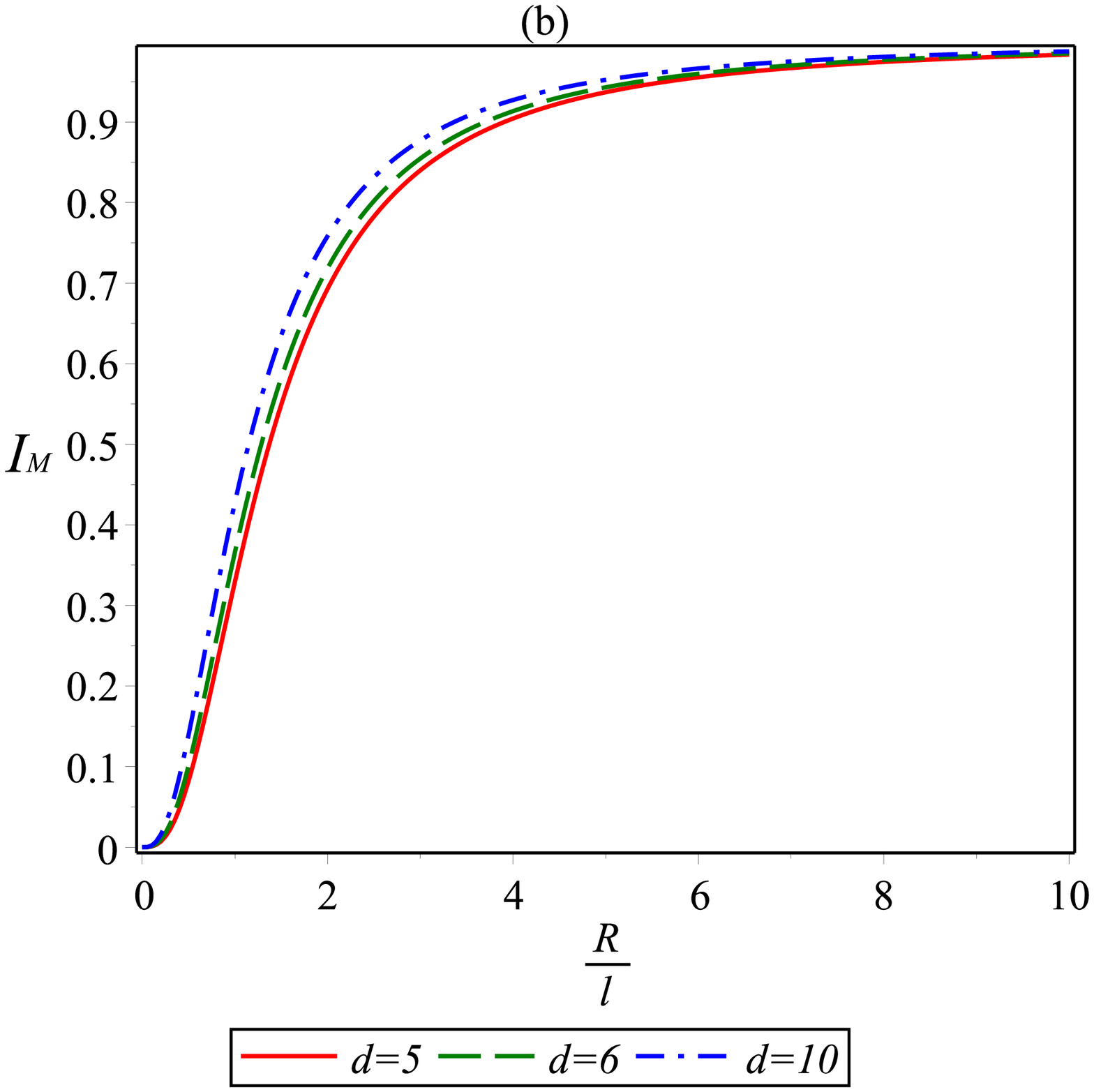}
\end{array}$
\end{center}
\caption{Moment of inertia and moment of inertia per unit mass in
terms of $R/l$: (a) $d=6$ and $l=1$; (b) $r_{0}/l=0.01$.} \label{figI}
\end{figure}

The moment of inertia of the thin black ring is
\begin{equation}\label{I}
I=\frac{\Omega_{d-3}r_{0}^{d-4}R^{4}}{8}\frac{\left(d-3+(d-2)\frac{R^{2}}{l^{2}}\right)^{\frac{3}{2}}}{\left(1+\frac{R^{2}}{l^{2}}\right)
\sqrt{1+(d-2)\frac{R^{2}}{l^{2}}}}.
\end{equation}
obtained from the classical relation $J=I\Omega^{2}$.
It is an increasing function of
$R$ and a decreasing function of $d$. We depict its typical
behavior  in Fig. \ref{figI} (a) for  $d=6$, while in
Fig. \ref{figI} (b) we plot
$I_{M}\equiv\frac{I}{M}$ and see that it becomes constant
at large values of $R$.  We can rewrite  \eqref{I} as
\begin{equation}\label{new I}
I=\mathcal{M}{\mathcal{R}}^{2},
\end{equation}
in order to have a match with the
classical relation by defining a new mass parameter
\begin{equation}\label{new M}
\mathcal{M}\equiv \frac{MR}{2}
\end{equation}
  in complete agreement with the mass given in
Ref. \cite{10},
where
\begin{equation}\label{new R}
\mathcal{R}=\sqrt{\frac{2}{d-2}}R\left(\frac{\left(d-3+(d-2)\frac{R^{2}}{l^{2}}\right)^{3}}{\left(1+\frac{R^{2}}{l^{2}}\right)^{5}
(1+(d-2)\frac{R^{2}}{l^{2}})}\right)^{\frac{1}{4}},
\end{equation}
is the reduced radius. For $d=4$ and $l\rightarrow\infty$ we see that $\mathcal{R}=R$.

\begin{figure}
\begin{center}$
\begin{array}{cccc}
\includegraphics[width=50 mm]{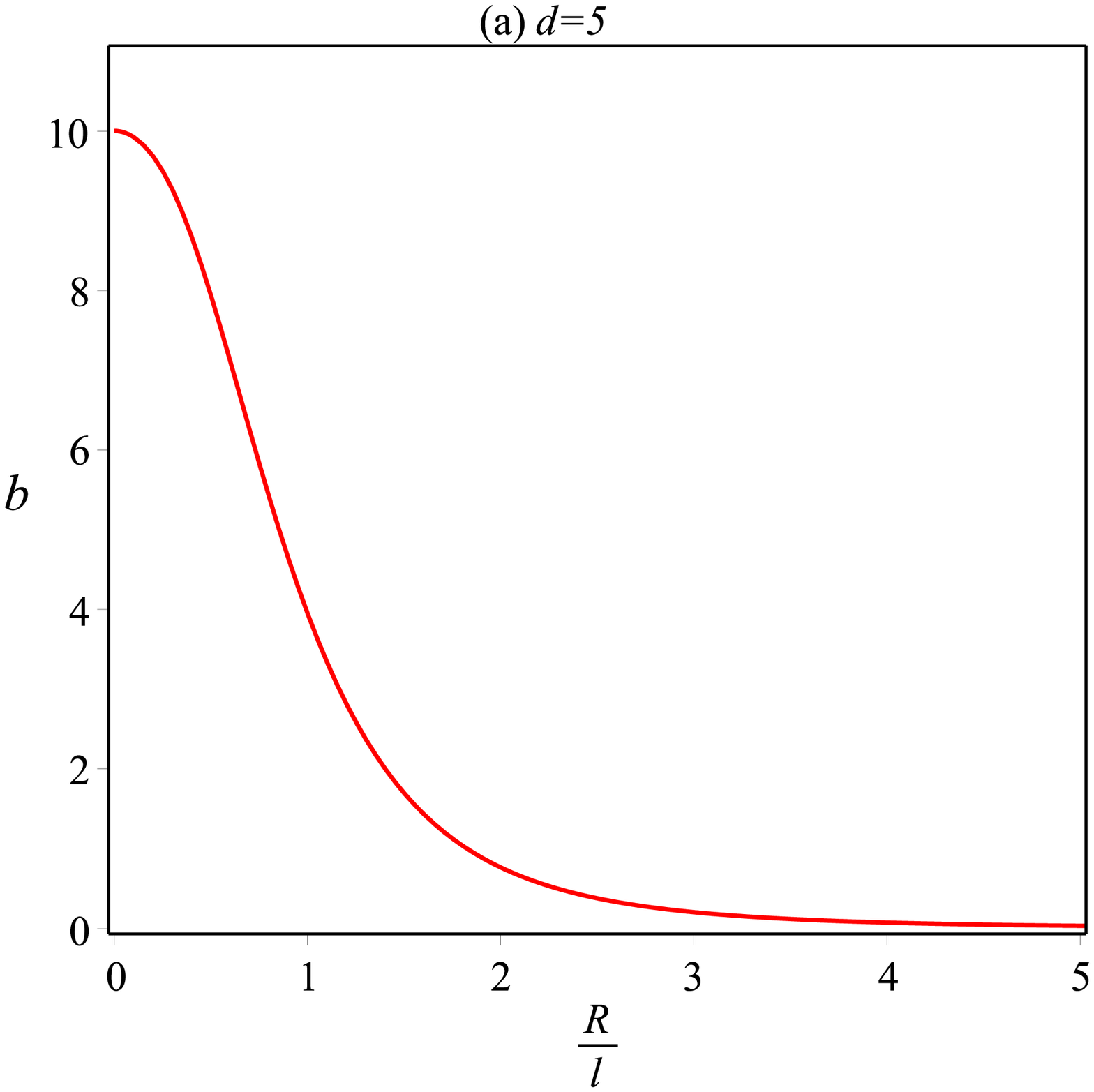}\includegraphics[width=50 mm]{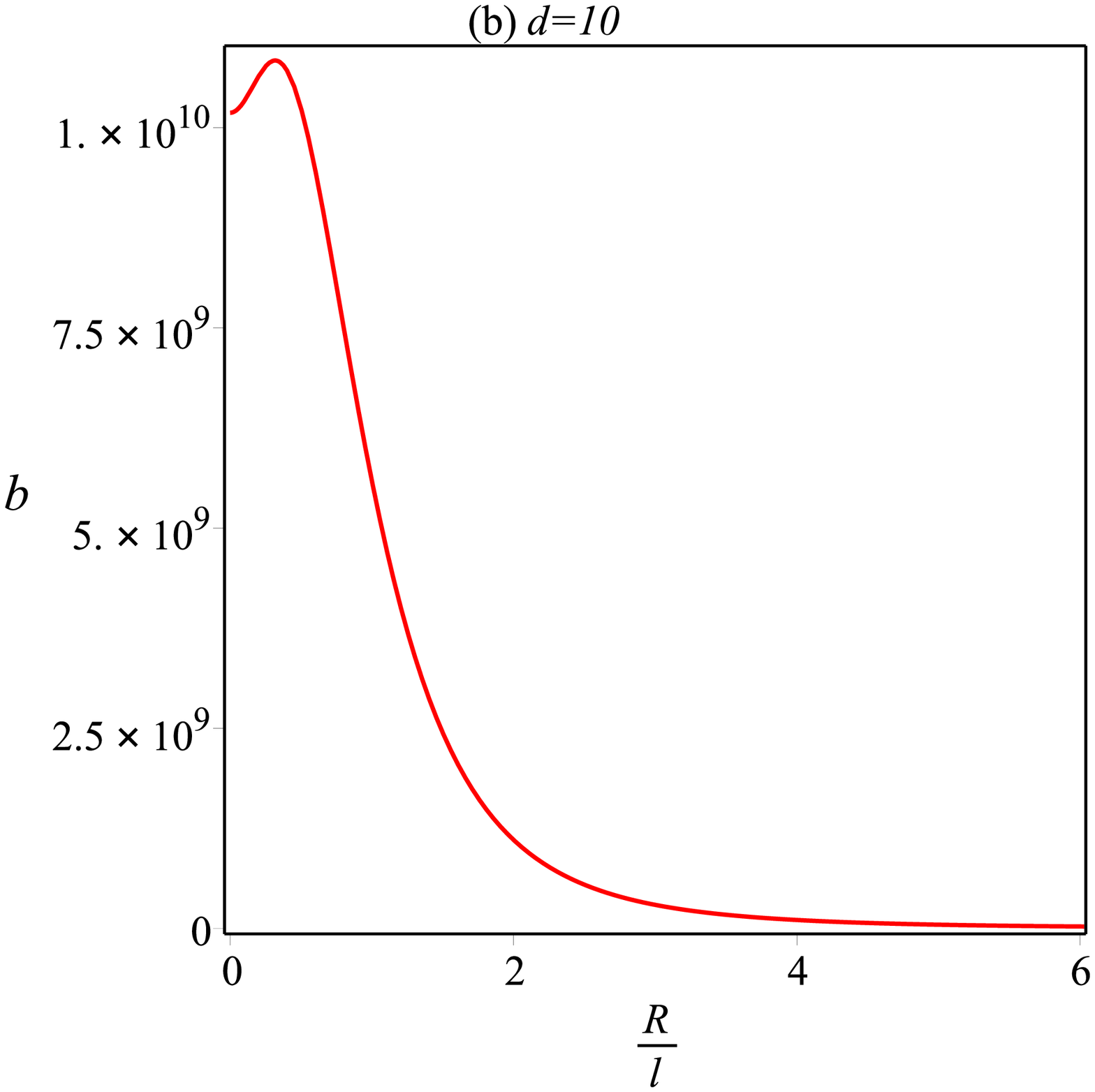}\includegraphics[width=50 mm]{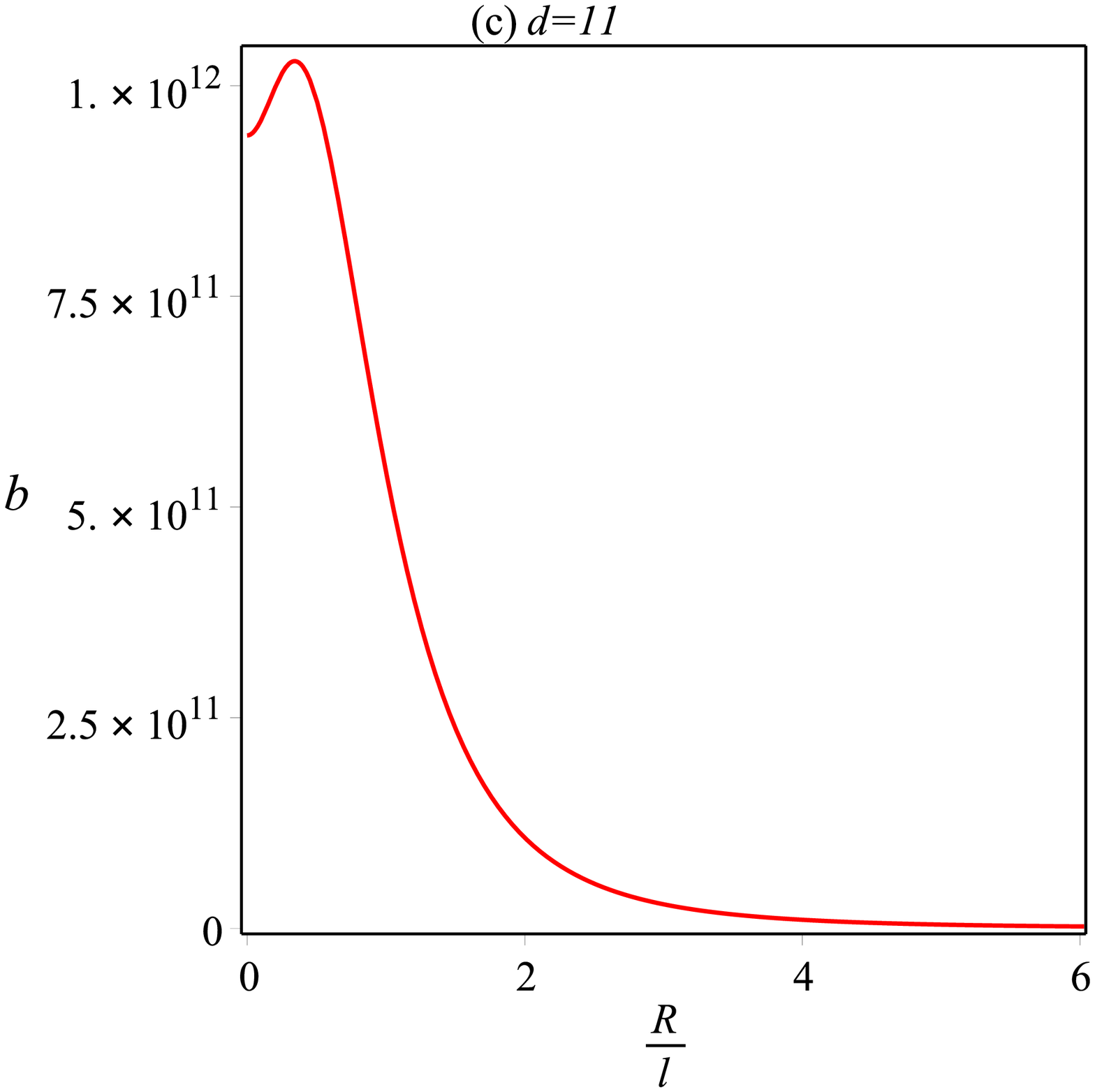}
\end{array}$
\end{center}
\caption{Typical behavior of the values of the angular momentum
per mass squared in terms of $R/l$ for $r_{0}/l=0.01$.}
\label{figb}
\end{figure}

Consider next the quantity
\begin{equation}
b = a/M =J/M^2.
\end{equation}
This is an interesting quantity that varies as $1/r_0^{d-4}$
(in Planck unit) that (for sufficiently large $R/l$) is generally a decreasing function of
$R$ and vanishes at large values of $R$.  It has a maximum at
\begin{equation}
\frac{R}{l} =  \sqrt{\frac{6-3d  + \sqrt{25 d^2 - 196 d + 388}}{8(d-2)}}
\end{equation}
provided $d \geq 7$; otherwise it is maximized at $R=0$.  We illustrate this
behavior  in  Fig. \ref{figb} for $d=5, 10, 11$ (see solid red lines).  As $d$ increases the maximum value of $b$ is a monotonically increasing function of $d$.
The value of $b$  in contrast to that of a $4D$ Kerr black hole where $-1<b<1$. Other curves with $\alpha=1$ related to the thermal fluctuations which described in next section.\\
Using (\ref{new M}), we find that the maximum for
$\bar{a}\equiv\frac{J}{\mathcal{M}}$ is at
\begin{equation}\label{new-a-max}
R({\bar{a}}_{max})=\sqrt{\frac{d^{2}-7d+13}{d-2}}l.
\end{equation}
Using it in ${\bar{a}}$ we find,
\begin{equation}\label{a-max}
{\bar{a}}_{max}={\frac {2\sqrt { \left( 1+\sqrt {{d}^{2}-7\,d+13} \right)  \left( d-
3+\sqrt {{d}^{2}-7\,d+13} \right) }}{(d-2)\left( 1+{\frac {\sqrt {{d}^
{2}-7\,d+13}}{d-2}} \right)^{3/2}} },
\end{equation}
so that ${\bar{a}}_{max}$ is a decreasing function of $d$
(independent of $l$) and we find for $5\leq d\leq11$ that  $1.075 \geq{\bar{a}}_{max}\geq 1.028$.

\section{Thermodynamics}

The thermodynamic quantities for a thin AdS black ring have been previously calculated  with
\begin{equation}\label{T}
T=\frac{(d-4)^{\frac{3}{2}}\sqrt{1+\frac{R^{2}}{l^{2}}}}{4\pi
r_{0}\sqrt{d-3+(d-2)\frac{R^{2}}{l^{2}}}}
\end{equation}
 being the temperature, computed from the surface gravity at the horizon and
\begin{equation}\label{S_{0}}
S_{0}=\frac{\pi}{2{l_{p}^{d-2}} }r_{0}^{d-3}\Omega_{d-3}R\sqrt{\frac{d-3+(d-2)\frac{R^{2}}{l^{2}}}{d-4}}.
\end{equation}
is the dimensionless entropy \cite{18}, where we have explicitly included the Planck area.

\begin{figure}
\begin{center}$
\begin{array}{cccc}
\includegraphics[width=55 mm]{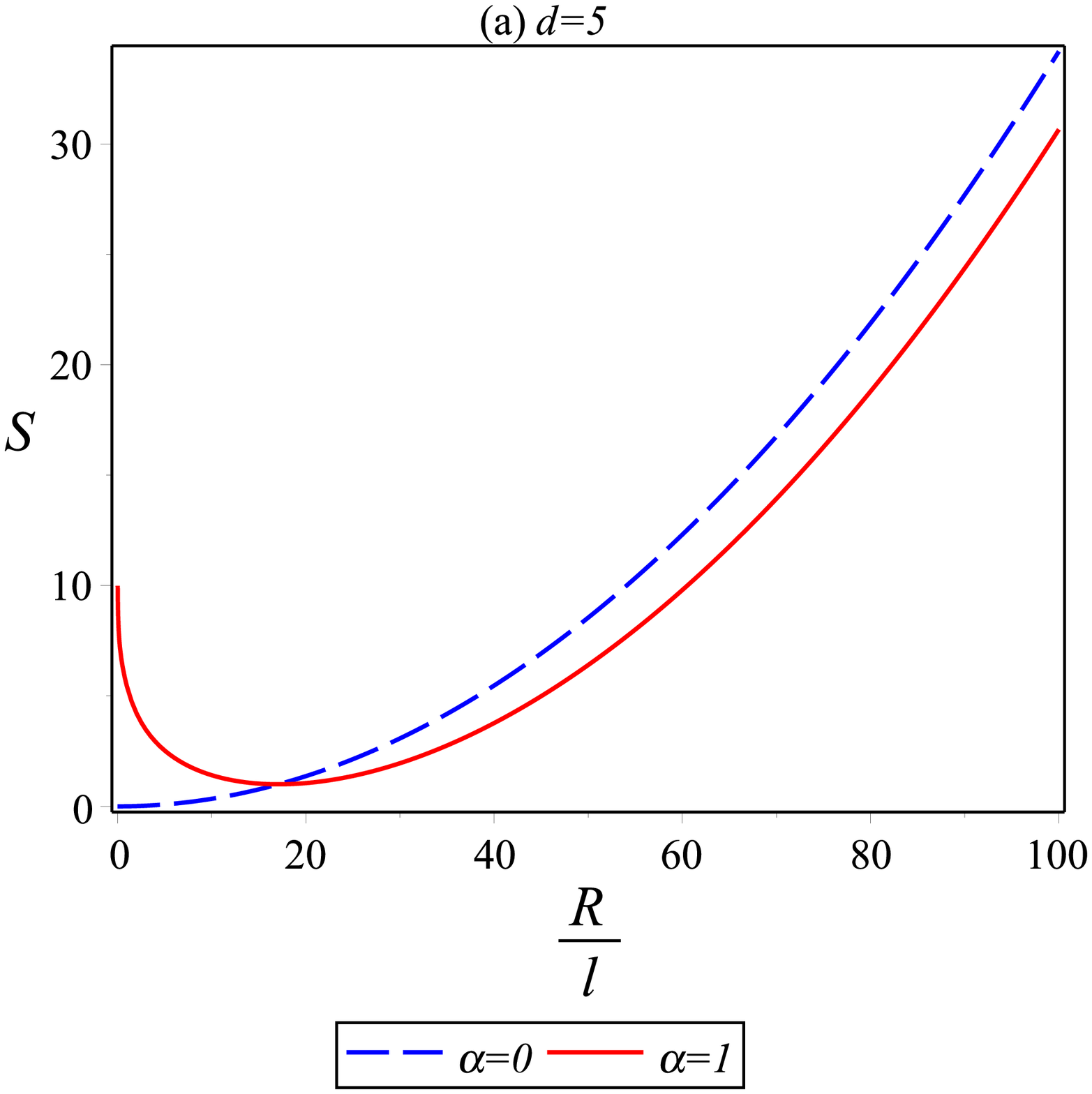}\includegraphics[width=55 mm]{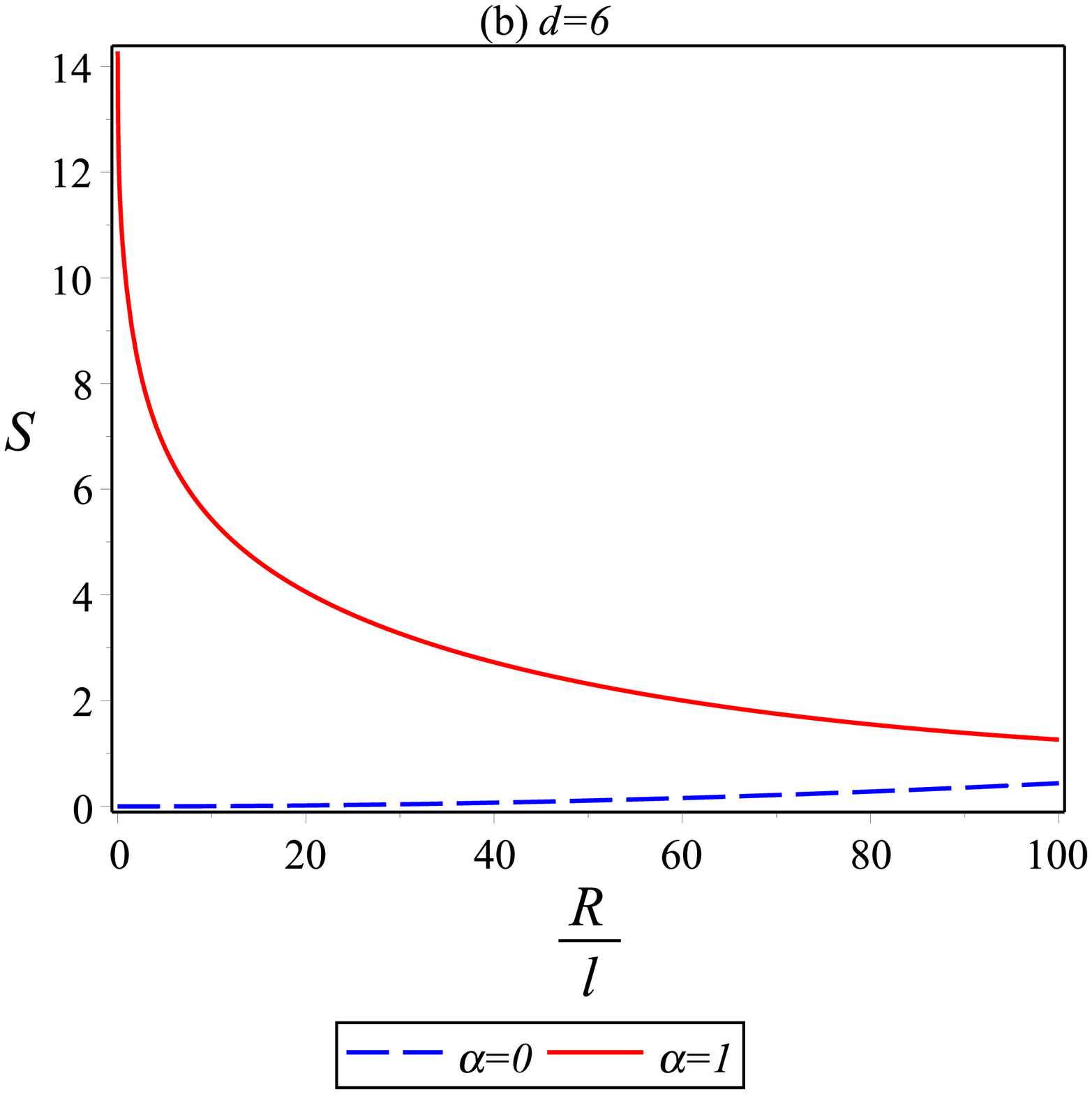}
\end{array}$
\end{center}
\caption{Typical behavior of the entropy \eqref{S_{0}} (dashed blue curve) and
logarithmic corrected entropy (solid red curve) in
terms of $R/l$ for $r_{0}/l=0.01$, and $l_{p}=1$.} \label{figS}
\end{figure}

It is clear that $T$ is a decreasing function of $R/l$. As the black ring radiates its mass decreases
and so it will reduce its size. However, other effects such as thermal (or quantum)
fluctuations can eventually become important.  Such corrections to lowest order yield to the entropy of black
objects as following \cite{Mann:1997hm,22},
\begin{equation}\label{S}
\bar{S}=S_{0}- \frac{\alpha}{2} \ln{\left( S_{0}\right)},
\end{equation}
where $\alpha$ is dimensionless correction parameter  parameterizing the effect of the logarithmic correction \cite{epl}. Above relation can be obtained by using Taylor expansion of partition function of canonical ensemble around the equilibrium temperature and neglecting higher order terms.
We depict the behavior of the logarithmically corrected entropy \eqref{S} in Fig. \ref{figS}. We see that effects
of thermal fluctuations are extremely significant at smaller
values of $R$.
In Fig. \ref{figS} (a),  corresponding to $d=5$, we
 see that there is a critical radius (denoted by $R_{c}$)  where $S=S_{0}$,
below which corrections due to thermal fluctuations dominate over the semiclassical value given in
\eqref{S_{0}}; in Fig. \ref{figS} (a)  the critical radius  $R_c/l=18$. It is clear that $S\rightarrow S_{0}$ at large $R/l$; indeed for
 $R = R_{c}$ and $R\gg R_{c}$ we have $S\approx S_{0}$ as well as $\alpha=0$.\\
The critical value $R_c$ is easily obtained from the vanishing of the logarithmic term in \eqref{S} as,
\begin{equation}\label{Rc}
\frac{R_{c}}{l}=\frac{\sqrt{2}}{2}\sqrt{\frac{\sqrt{16r_{0}^{6}l_{p}^{2d}(d-4)\left[\Gamma\left(\frac{d}{2}\right)\right]^{2}
+\pi^{d}r_{0}^{2d}l^{2}l_{p}^{4}(d-2)(d-3)^{2}}}{r_{0}^{d}\pi^{\frac{d}{2}}l l_{p}^{2}(d-2)^{\frac{3}{2}}}-\frac{d-3}{d-2}},
\end{equation}
which is clearly a decreasing function of $r_{0}$. At sufficiently small $r_0$ this will violate the approximations
used to obtain the thermodynamic parameters for the black ring. In Fig. \ref{figS} it is given by
the intersection of the solid red (presence of thermal fluctuation) and the dashed blue
(absence of thermal fluctuation) lines.

\subsection{Stability}
\begin{figure}
\begin{center}$
\begin{array}{cccc}
\includegraphics[width=55 mm]{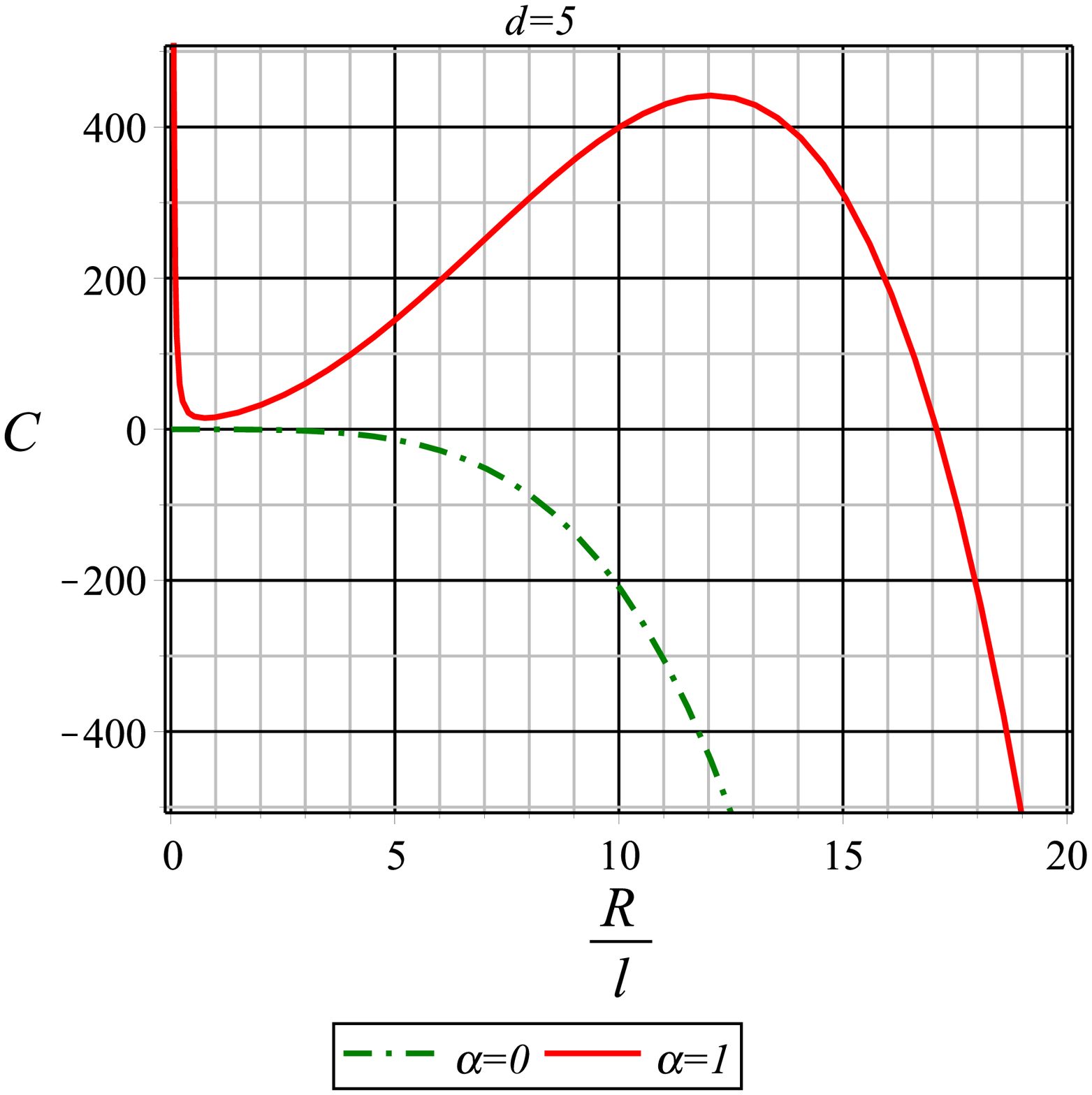}\includegraphics[width=55 mm]{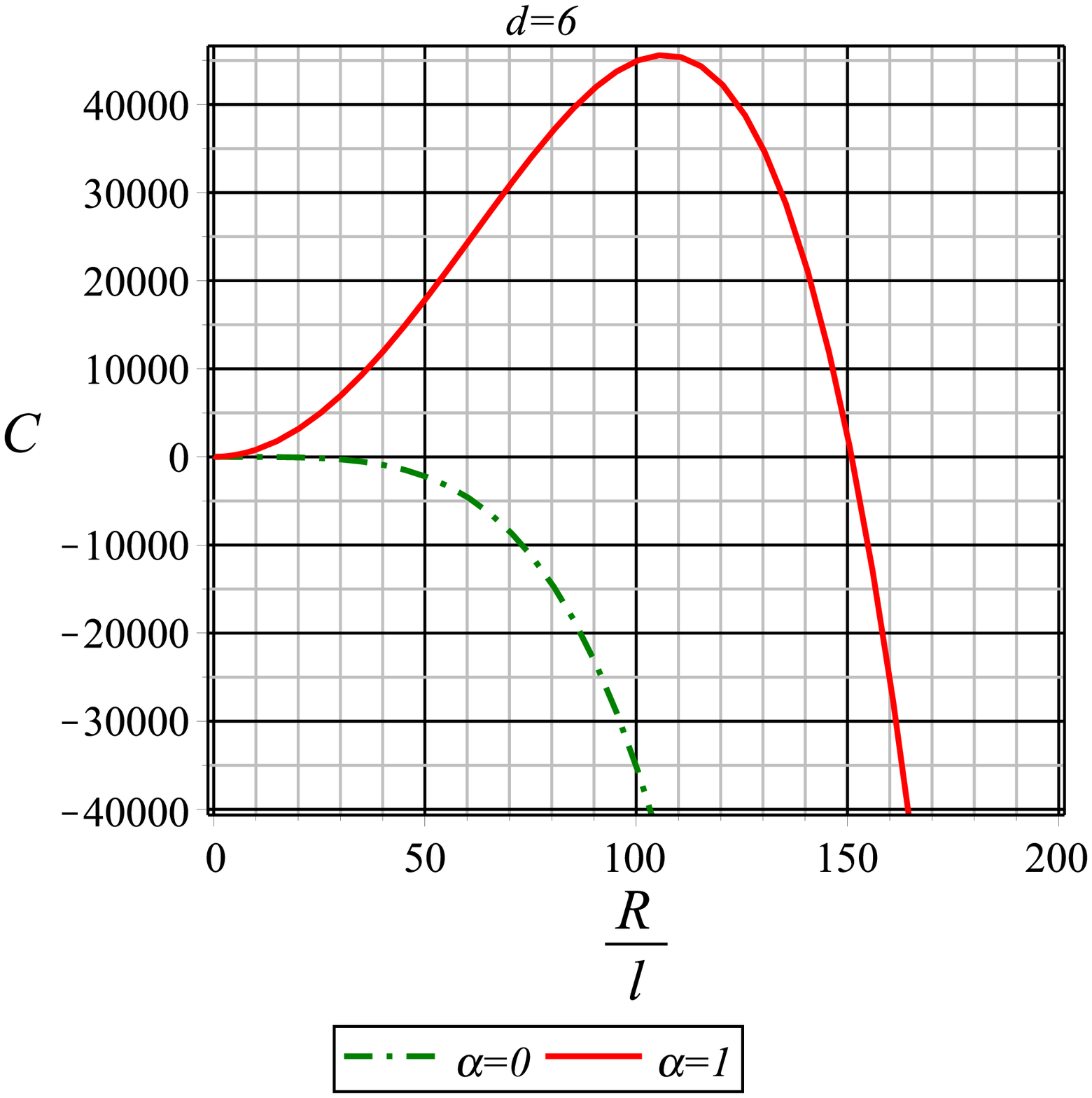}\\
\includegraphics[width=55 mm]{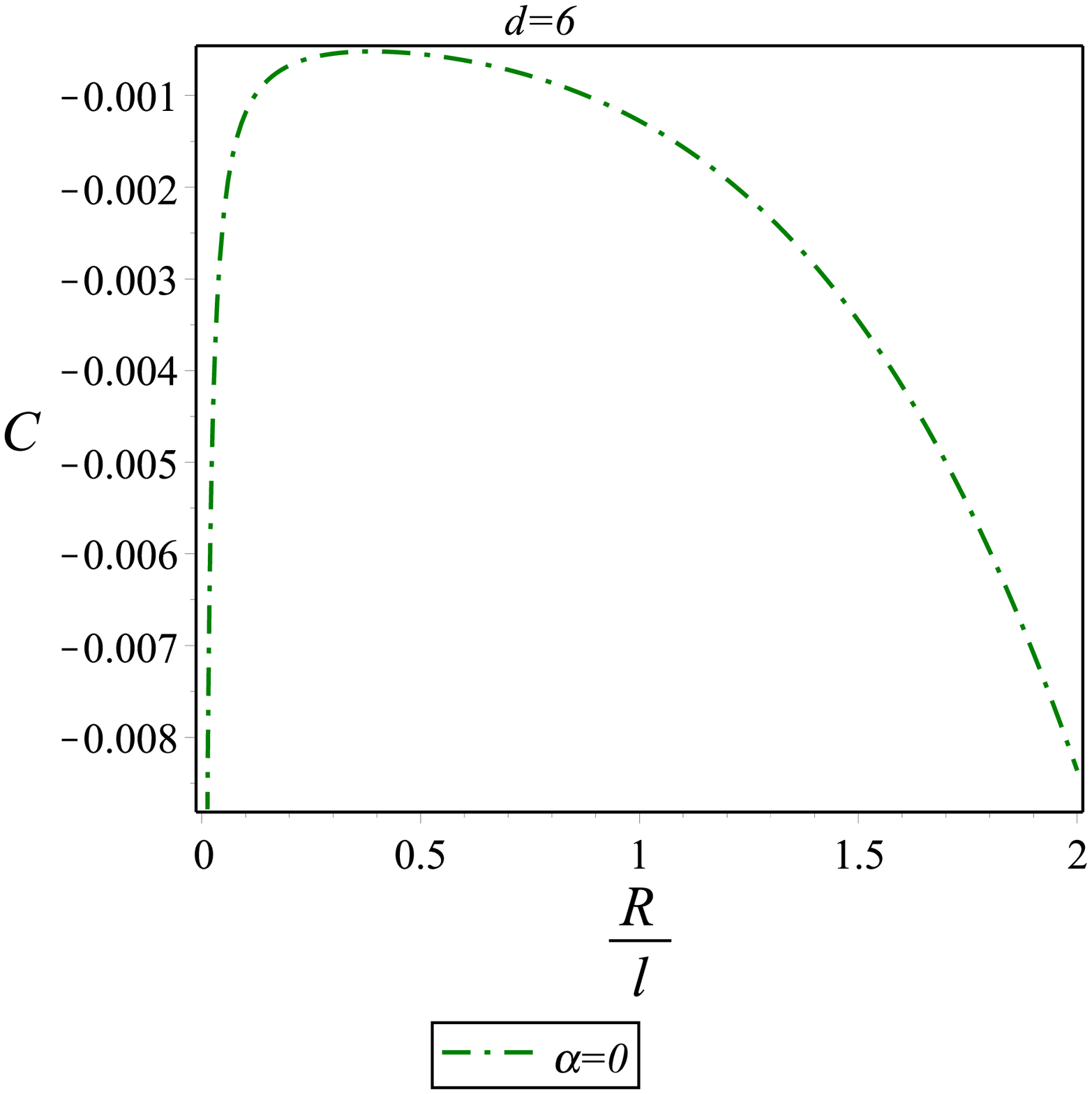}\includegraphics[width=55 mm]{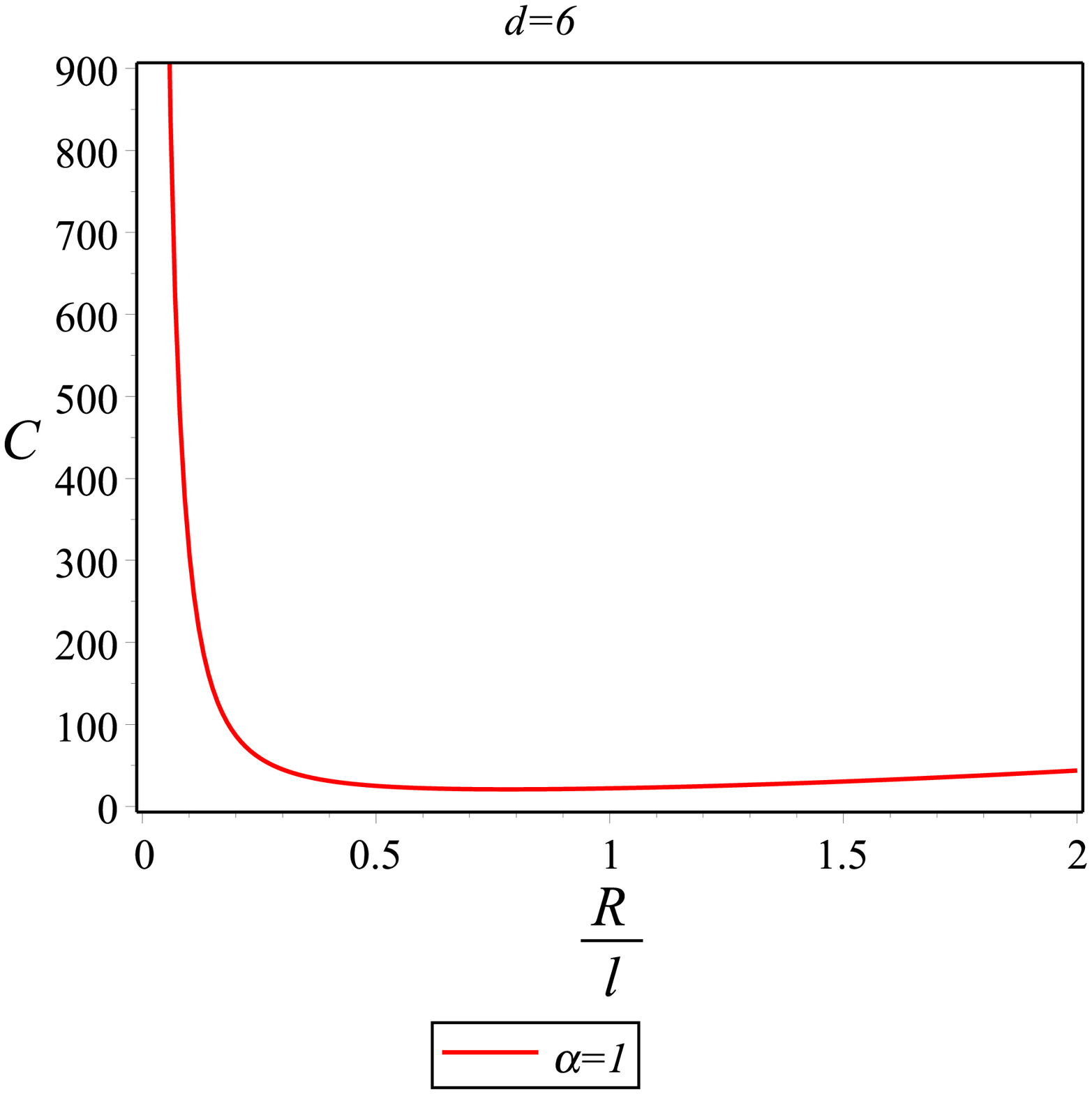}
\end{array}$
\end{center}
\caption{Typical behavior of the specific heat in terms of $R/l$
for $r_{0}/l=0.01$ and $l_{p}=1$. Lower plots are the same as
upper right to show the behavior of specific heat near the
origin.} \label{figC}
\end{figure}

By using the equation \eqref{S} one can obtain the specific heat as
\begin{equation}\label{C}
C=T(\frac{d\bar{S}}{dT}) = T\frac{d\bar{S}}{dR} \left(\frac{dT}{dR}\right)^{-1}=C_{0}+\alpha C_{1},
\end{equation}
where
\begin{eqnarray}\label{C2}
C_{0}&=&T\left(\frac{dS_{0}}{dT}\right)\nonumber\\
&=&-\frac{\pi^{d/2}r_{0}^{d-3}(l^{2}+R^{2})\left((d-3)l^{2}+2(d-2)R^{2}\right)\sqrt{\left((d-3)l^{2}+(d-2)R^{2}\right)}}
{\Gamma\left(\frac{d-2}{2}\right)l^{3}Rl_{p}^{d-2}\sqrt{(d-4)}},
\end{eqnarray}
is the semiclassical specific heat, while
\begin{equation}\label{C3}
C_{1}=\frac{(l^{2}+R^{2})((d-3)l^{2}+2(d-2)R^{2})}{2l^{2}R^{2}},
\end{equation}
is the correction term. It is clear that the only divergent point is $R=0$. In the case of $\alpha=0$ we have $C_{0}\rightarrow-\infty$ at $R=0$, while $C_{1}\rightarrow\infty$ at origin. Mentioned divergencies are clear from lower plots of Fig. \ref{figC} in $d=6$ (for example) which are drawn for the near origin behavior.\\

The sign  of the specific heat determines the thermodynamic
stability of the system. The system is in a stable phase if $C\geq0$
and phase transitions may occur at the asymptotic points. In Fig. \ref{figC} we plot the specific heat for $d=5$ and
$d=6$ as a function of $R/l$; other dimensions
yield similar results. Dash dotted green lines of Fig. \ref{figC} show the plot of the
uncorrected specific heat (equation (\ref{C2})). From the figure it is
clear that the thin black ring in AdS space-time is completely
unstable because of the negativity of its specific heat.
However, the  corrected specific heat, obtained using \eqref{S} in \eqref{C}, is  positive for sufficiently small $R/l$, and asymptotes
to its classical value at large $R/l$. The
thin black ring in AdS space-time may be therefore be thermodynamically stable once quantum corrections and/or thermal fluctuations
are taken into account.
Similar results hold for any $d$.\\
The thermodynamic volume (specific volume) of a thin
AdS black ring, is given by,
\begin{equation}\label{V}
V=\Omega_{d-3}\frac{\pi r_{0}^{d-4}}{(d-1)l_{p}^{d-2}}R^{3}\sqrt{1+\frac{R^{2}}{l^{2}}},
\end{equation}
while the thermodynamic pressure is
\begin{equation}\label{P0}
P=\frac{(d-1)(d-2)}{16\pi l^{2}}.
\end{equation}
Consider next the Gibbs free energy which is given by,
\begin{equation}\label{G}
G=\bar{M}-T\bar{S},
\end{equation}
where $\bar{M}$ is the corrected mass formula. In order to calculate it and hence Gibbs free energy, we need the modified first law of thermodynamics which studied in next subsection.

\subsection{The First law of Thermodynamics}

It is easy to check that the first law of thermodynamics is
satisfied in case of $\alpha=0$,
\begin{equation}\label{fr0}
dM=TdS_{0}+\Omega dJ+VdP,
\end{equation}
The Smarr relation also holds as below
\begin{equation}\label{Smarr0}
(d-3)M=(d-2)TS_{0}+(d-2)\Omega J-2PV.
\end{equation}
The first law will be correspondingly modified.  We can write
\begin{eqnarray}\label{MJ}
\bar{M}&=&M+\alpha  M_1, \nonumber\\
\bar{J}&=&J+\alpha  J_1, \nonumber\\
\bar{V}&=&V+\alpha  V_1,
\end{eqnarray}
where $M$, $J$, and $V$ are respectively given by equations \eqref{M}, \eqref{J} and  \eqref{V}.
The modified first law of thermodynamics
\begin{equation}\label{fr00}
d\bar{M}=Td\bar{S}+\Omega d\bar{J}+\bar{V}dP
\end{equation}
is satisfied if
\begin{equation}\label{fr1}
d M_{1}=TdS_{1}+\Omega d J_{1}+V_{1}dP,
\end{equation}
where $S_{1}=-\frac{\ln{S_{0}}}{2}$. It is the fact that the thin black ring solution depends on three independent parameters $R$ and $l$ and $r_0$, and so the modified first law (\ref{fr00}) leads to a set of coupled partial differential equations for the corrections. This gives
\begin{align}
\frac{\partial M_1}{\partial R} &= T  \frac{\partial S_1}{\partial R} + \Omega  \frac{\partial J_1}{\partial R} ,  \label{M1-1}\\
\frac{\partial M_1}{\partial l} &= T  \frac{\partial S_1}{\partial l} + \Omega  \frac{\partial J_1}{\partial l}  - V_1  \frac{(d-1)(d-2)}{8\pi l^{3}}, \label{V1-1}\\
\frac{\partial M_1}{\partial r_{0}} &= T  \frac{\partial S_1}{\partial r_{0}} + \Omega  \frac{\partial J_1}{\partial r_{0}}   \label{J1-1}
\end{align}
and, noting the $r_0$ dependence of $T$ and $S_1$ using \eqref{S},  this last equation implies that both $M_1$ and $J_1$
have the form $A(R,l)/r_0 +B(R,l)$, where $A$ and $B$ are determined from the other two equations.  Remarkably,
one can exactly integrate \eqref{M1-1} to obtain
\begin{align}
M_1  &= (d - 2) \sqrt{(d - 4)^3} \frac{(1+ \frac{R^2}{l^2})^{3/2} }{8\pi r_0 \sqrt{(d - 3)+ (d - 2)\frac{R^2}{l^2}} }  + \hat{M}_1(R,l) \label{M1-sol}\\
J_1  &=  R \sqrt{(d - 4)^3} \frac{\sqrt{ 1 + (d - 2)\frac{R^2}{l^2}} }{8\pi r_0 }  + \hat{J}_1(R,l) \label{J1-sol}\\
V_1&= \frac{R^2 \sqrt{(d - 4)^3}  \sqrt{1+ \frac{R^2}{l^2}}}{r_0 (d - 1) \sqrt{(d - 3)+ (d - 2)\frac{R^2}{l^2}} } + \hat{V}_1(R,l)  \label{V1-sol}
 \end{align}
where
 \begin{align}
\frac{\partial \hat{M}_1 }{\partial R} &=   \Omega  \frac{\partial \hat{J}_1}{\partial R} \label{M1hat-eq}\\
\frac{\partial \hat{M}_1}{\partial l} &=  \Omega  \frac{\partial \hat{J}_1}{\partial l}  - \hat{V}_1(R,l)  \frac{(d-1)(d-2)}{8\pi l^{3}}, \label{V1-hat}
 \end{align}
must be satisfied.

There is considerable ambiguity in obtaining the functions $(\hat{M}_1, \hat{J}_1,\hat{V}_1)$, since the preceding set of equations is an underdetermined system.
The simplest solution is to assume each of $(\hat{M}_1, \hat{J}_1,\hat{V}_1)$ to zero, in which case
\begin{eqnarray}\label{MJ}
\bar{M}&=& \frac{r_{0}^{d-4}\Omega_{d-3}R(d-2)\left(1+\frac{R^2}{l^2}\right)^{\frac{3}{2}}}{8l_{p}^{d-2}} +\alpha  (d - 2)  \frac{\sqrt{(d - 4)^3(1+ \frac{R^2}{l^2})^{3}} }{8\pi r_0 \sqrt{(d - 3)+ (d - 2)\frac{R^2}{l^2}} }   \nonumber\\
\bar{J}&=& \frac{r_{0}^{d-4}\Omega_{d-3}R^{2}\sqrt{(1+(d-2)\frac{R^{2}}{l^{2}})(d-3+(d-2)\frac{R^{2}}{l^{2}})}}{8l_{p}^{2}} +\alpha  R \sqrt{(d - 4)^3} \frac{\sqrt{ 1 + (d - 2)\frac{R^2}{l^2}} }{8\pi r_0 } , \nonumber\\
\bar{V}&=&\Omega_{d-3}\frac{\pi r_{0}^{d-4}}{(d-1)l_{p}^{d-2}}R^{3}\sqrt{1+\frac{R^{2}}{l^{2}}}+\alpha  \frac{R^2 \sqrt{(d - 4)^3}  \sqrt{1+ \frac{R^2}{l^2}}}{(d - 1) r_0\sqrt{(d - 3)+ (d - 2)\frac{R^2}{l^2}} }
\end{eqnarray}
are the corrected thermodynamic parameters of the black ring to order $\alpha$. These quantities satisfy the first law \eqref{fr00}. with the thermodynamic parameters replaced with their barred counterparts.
Note that this choice of integration constants ensures that the corrections to $J$ and $V$ vanish as $R\to 0$, but that the correction to $M$ does not.
Furthermore the Smarr relation   \eqref{Smarr0} is not satisfied.  This is expected -- the Smarr relation follows from Eulerian scaling of the thermodynamic variables
but the correction \eqref{S} to the entropy does not have the same scaling properties as $S_0$.

\begin{figure}
\begin{center}$
\begin{array}{cccc}
\includegraphics[width=60 mm]{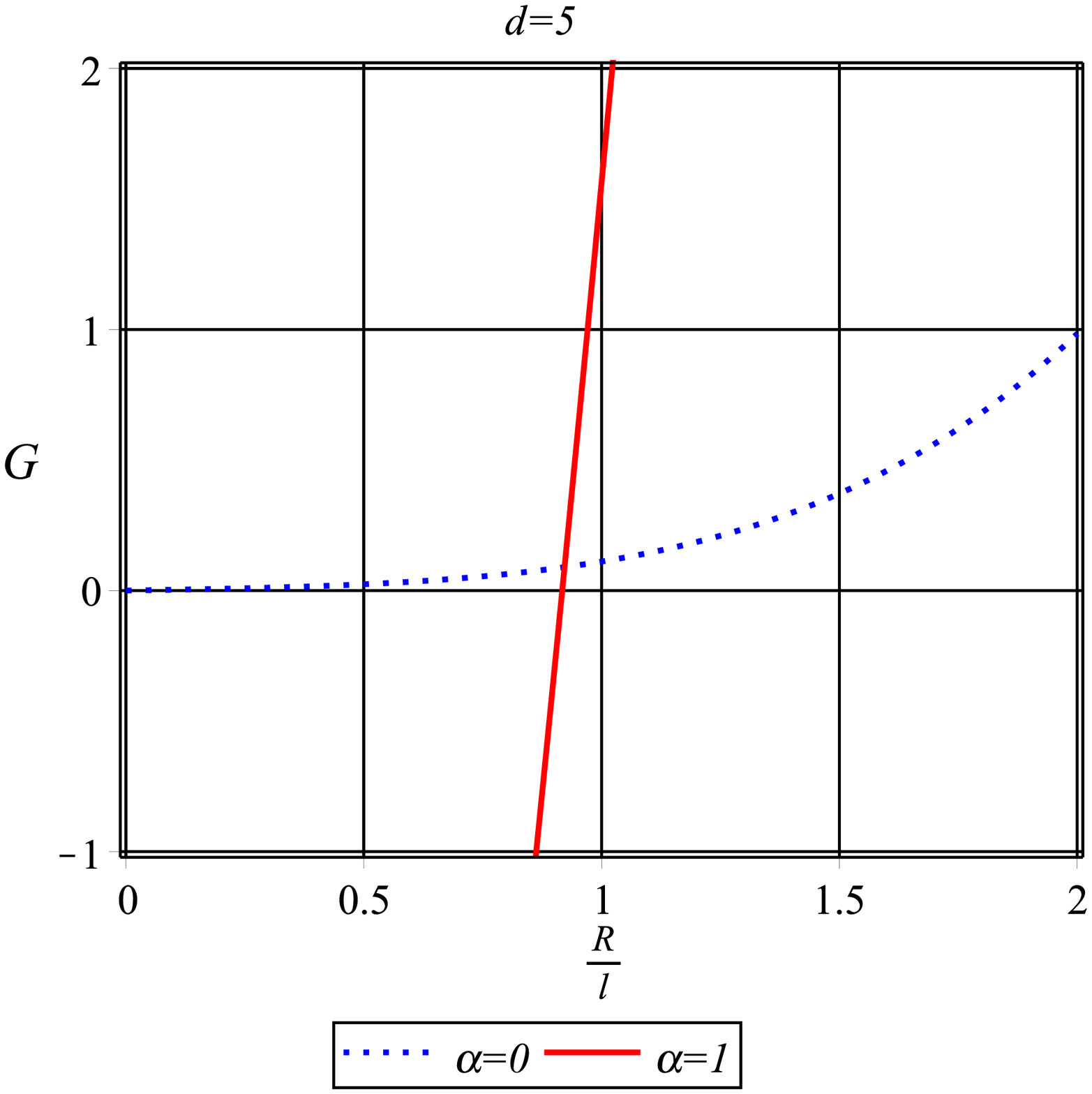}\includegraphics[width=60 mm]{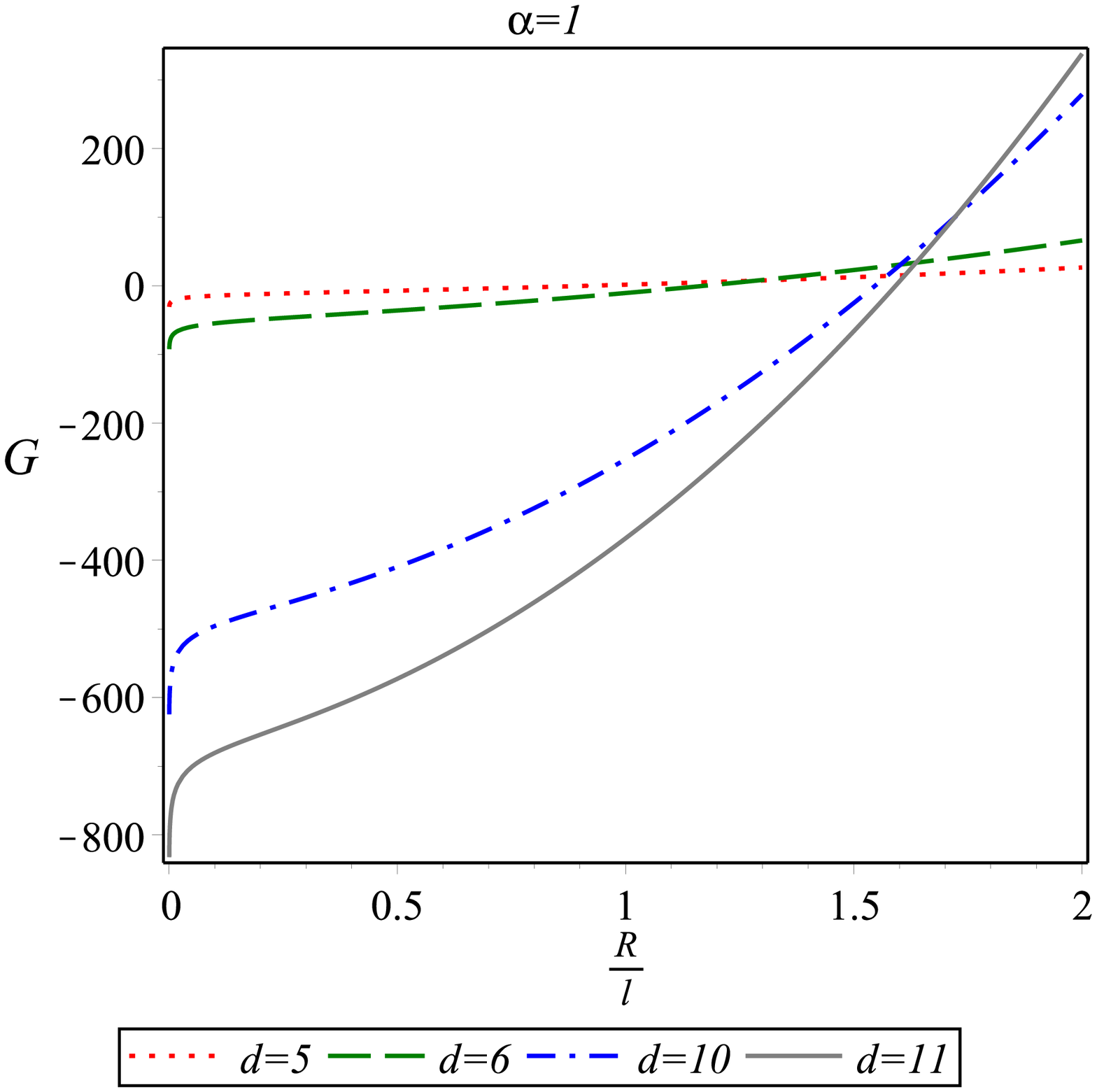}
\end{array}$
\end{center}
\caption{Typical behavior of the Gibbs free energy in terms of
$R/l$ for $r_{0}/l=0.01$ and $l/l_{p}=1$.} \label{figG}
\end{figure}

We compute the Gibbs free energy via the equation (\ref{G}), obtaining
\begin{equation}\label{GG}
G=G_{0}+\alpha G_{1},
\end{equation}
where $G_{0}$ is value of Gibbs free energy at $\alpha=0$, given by
\begin{eqnarray}\label{G0}
G_{0}&=& {\frac {r_{0}^{d-4}{\pi }^{d/2-1}R }{4\Gamma  \left( d/2-1 \right) l_{p}^{d-2}} \left( 1+
{\frac {{R}^{2}}{{l}^{2}}} \right) ^{1/2}}
\left(\frac { \left( d-2 \right) {R}^{2}}{{l}^{2}} +2 \right)
\end{eqnarray}
(see dotted blue lines of Fig. \ref{figG}), and $G_{1}$ is the leading correction, given by
\begin{eqnarray}\label{G1}
 G_{1}&=&  {\frac {d-2}{8\pi \,r_{0}}\sqrt {{ \left( d-4 \right) ^{3} \left( 1+
{\frac {{R}^{2}}{{l}^{2}}} \right) ^{3} \left( d-3+{\frac { \left( d-2
 \right) {R}^{2}}{{l}^{2}}} \right) ^{-1}}}}\\
 &-&
 {\frac { \left( d-4 \right)^{3/2}}{8\pi r_{0}}\ln  \left( {\frac {
\pi r_{0}^{d-3}{\pi }^{d/2-1}R}{\Gamma  \left( d/2-1 \right) l_{p}^{d-2}
}\sqrt {{\frac {1}{d-4} \left( d-3+{\frac { \left( d-2 \right) {R}^{2}
}{{l}^{2}}} \right) }}} \right) {
\frac {\sqrt {1+{\frac {{R}^{2}}{{l}^{2}}}}}{\sqrt {d-3+{\frac { \left( d-2 \right) {R}^{2}}{{l}^{2}}}}}}
}\nonumber
\end{eqnarray}

We depict the behaviour of  $G$ in Fig. \ref{figG}.  The effect of nonzero $\alpha$ is to push the free energy curve to negative values for small $R/l$ and to larger positive values for large $R/l$ relative to $G_0$ as shown in the left panel of  Fig. \ref{figG}.   The intercept with the $R/l$ axis moves to the right as $\alpha$ increases, but saturates at a specific value  once $\alpha$ becomes sufficiently large. This intercept shifts rightward and the curve becomes steeper as the other parameters montonically change. We have illustrated the trend for increasing $d$ in the right panel of  Fig. \ref{figG}.  The same trend happens for  increasing $l$ and decreasing $r_0$ for any fixed $l_p$. Note that the derivation of the black ring is for sufficiently large $R/r_0$, and so these curves cannot be trusted for
small $R/l$.

\begin{figure}
\begin{center}$
\begin{array}{cccc}
\includegraphics[width=50 mm]{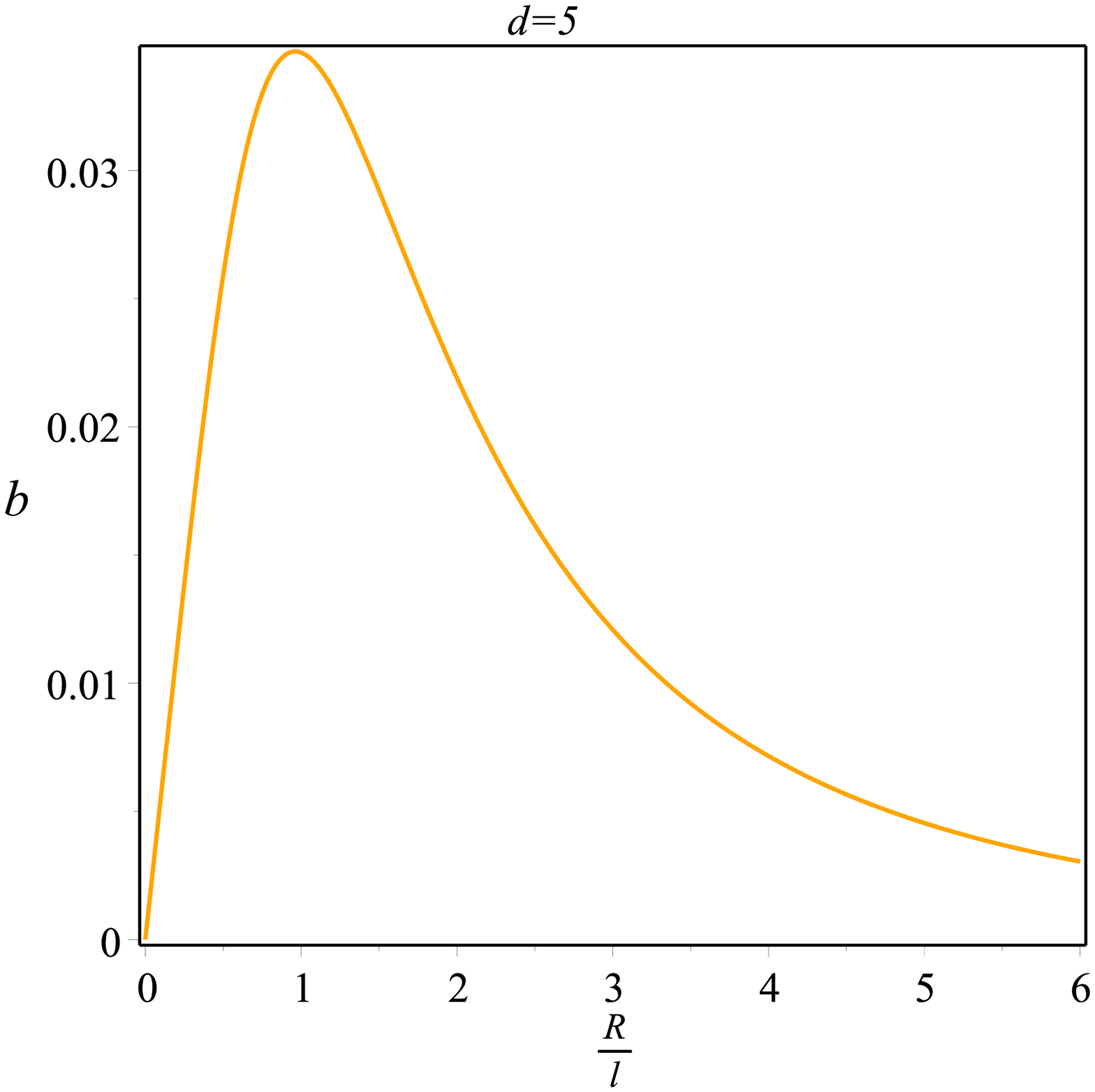}\includegraphics[width=50 mm]{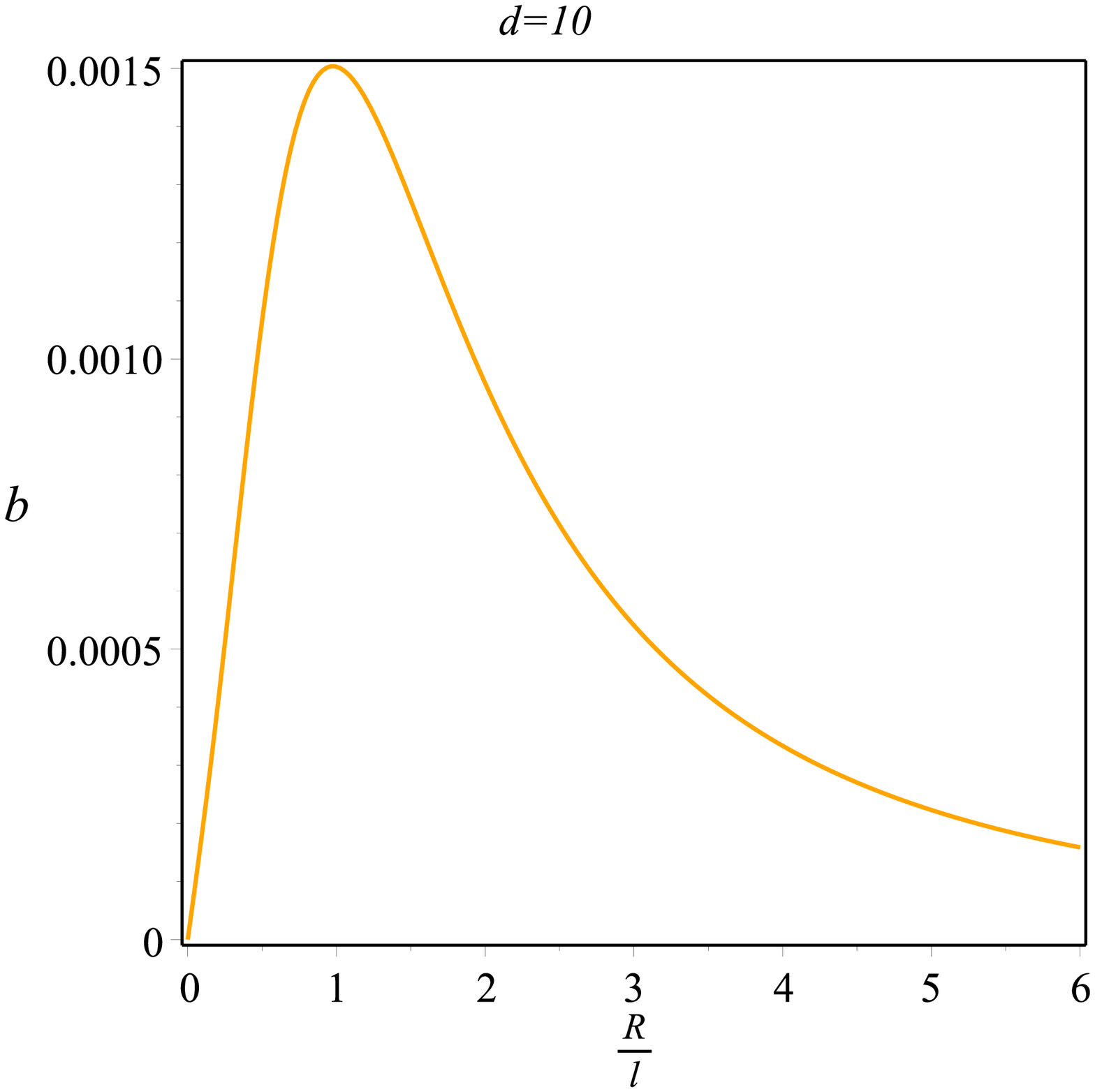}\includegraphics[width=50 mm]{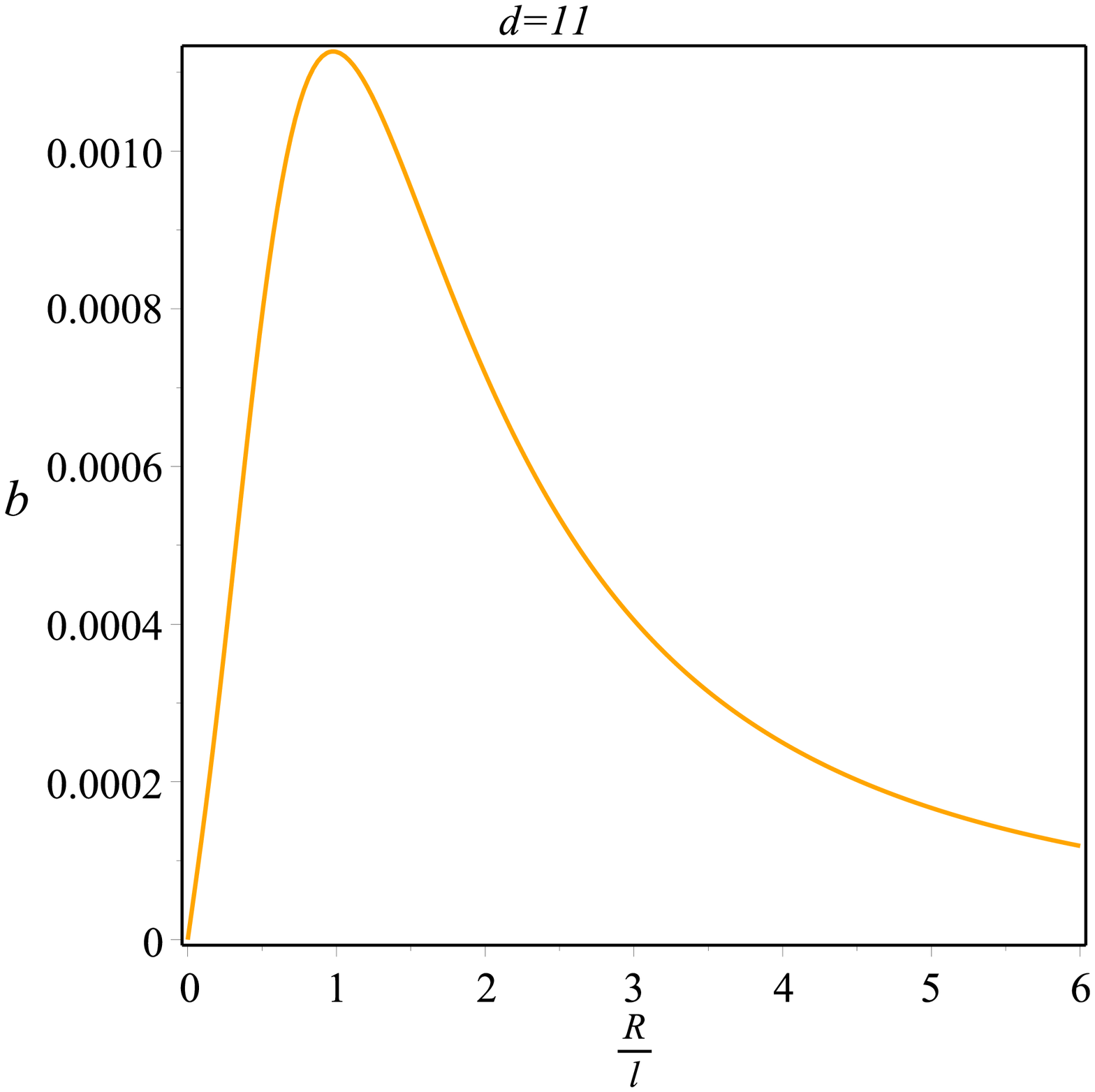}
\end{array}$
\end{center}
\caption{Typical behavior of the values of the angular momentum
per mass squared in presence of logarithmic correction, in terms of $R/l$ for $r_{0}/l=0.01$, and $\alpha=1$.}
\label{figba}
\end{figure}

The $b$ parameter will become
\begin{equation}\label{mod b}
b=\frac{\bar{J}}{{\bar{M}}^{2}}
\end{equation}
shown in Fig. \ref{figba}.
Comparing with Fig. \ref{figb}, we can see that the values of $b$ parameter is reduced due to the logarithmic correction, specially at small $R/l$.

These computations indicate that it is possible to systematically compute $\alpha$-dependent corrections to the various thermodynamic parameters of the black ring under the assumption that the intensive parameters ($T$, $P$, $\Omega$) remain unchanged.  It is possible, of course, to make different assumptions about these parameters, in which case the above analysis would need to be revisited.

\section{Discussions and Conclusion}
We have shown that a small thin AdS black ring may be
thermodynamically stable once thermal (or quantum) fluctuations
are taken into account. Such fluctuations correct the entropy of
black objects with a logarithmic term \cite{Mann:1997hm,22}, given
in equation \eqref{S}. These corrections become significant for small
values of the black ring radius (which corresponds to quantum
scales). It means that there is a critical radius $R_{c}$ below
which there are significant departures from standard
semi-classical behavior. As $R/l\rightarrow 0$ we observe a spike
in the modified entropy (given by the red bold line in Fig. \ref{figS}
(a)), with the trajectory eventually diverging at the $R/l=0$,
which may yield to evaporation of the black ring. Below the
critical radius the modified entropy of the black ring is a
decreasing function of $R/l$. As the number of space-time
dimensions increase, the critical radius becomes larger, but the
qualitative features almost remain the same.

To probe the thermodynamic stability of the ring, we performed a
specific heat analysis. We found that the modified specific heat
is positive at sufficiently small $R/l$, whereas the uncorrected
specific heat is always negative. Moreover, the trajectory for the
modified specific heat spikes near $R/l=0$ towards the positive
end, before diverging to positive infinity (see Fig.\ref{figC}).
This clearly suggests stability of the ring at quantum scales.

These results are corroborated by an analysis of the thermodynamic
potentials.  We find that the $alpha$-dependent corrections to
the other thermodynamic parameters can be computed such that the  first law of thermodynamics can be
satisfied \cite{PLB, PLB2}. The situation of the thin black ring solution is similar to the Kerr-AdS solution, which depends on three parameters; mass, angular momentum and AdS curvature. The uncorrected first law is satisfied for independent variations of the three parameters $R$, $l$, and $r_0$. Similarly, these three parameters should be varied in the corrected first law, leading to a set of three coupled partial differential equations.
The evolution of the Gibbs free energy indicates that small black rings with low angular momentum are
thermodynamically stable, whereas larger black rings of small angular momentum can undergo a Hawking-Page phase transition provided the angular momentum
of the ring can be dispersed into the radiation of thermal AdS.

We end by acknowledging the fact that stability of black
rings is a very delicate issue and it is difficult to cover all
the aspects of this topic in a single study. A more exhaustive and
thorough stability analysis taking all sorts of fluctuation
effects into account would be of interest and may be considered in
a future work.

\section*{Acknowledgments}

B.P. would like to thank Iran Science Elites Federation, Tehran,
Iran. P.R. acknowledges the Inter University Centre for Astronomy
and Astrophysics (IUCAA), Pune, India for granting visiting
associateship.  This work was supported in part by the Natural
Sciences and Engineering Research Council of Canada.


\end{document}